\documentclass[final,5p,times,twocolumn,sort&compress]{elsarticle}

\usepackage{amssymb}
\makeatletter
\newsavebox\myboxA
\newsavebox\myboxB
\newlength\mylenA

\newcommand*\xoverline[2][0.65]{%
    \sbox{\myboxA}{$\m@th#2$}%
    \setbox\myboxB\null
    \ht\myboxB=\ht\myboxA%
    \dp\myboxB=\dp\myboxA%
    \wd\myboxB=#1\wd\myboxA
    \sbox\myboxB{$\m@th\overline{\copy\myboxB}$}
    \setlength\mylenA{\the\wd\myboxA}
    \addtolength\mylenA{-\the\wd\myboxB}%
    \ifdim\wd\myboxB<\wd\myboxA%
       \rlap{\hskip 0.5\mylenA\usebox\myboxB}{\usebox\myboxA}%
    \else
        \hskip -0.5\mylenA\rlap{\usebox\myboxA}{\hskip 0.5\mylenA\usebox\myboxB}%
    \fi}
\makeatother

\journal{Chemical Physics Letters}

\begin{document}

\begin{frontmatter}



\title{Nanoscale heterogeneity at the aqueous electrolyte-electrode interface}


\author[label1]{David T. Limmer}
\address[label1]{Princeton Center for Theoretical Science, Princeton, NJ 08540}
\author[label2]{and Adam P. Willard} 
\address[label2]{Department of Chemistry, Massachusetts Institute of Technology, Cambridge, Massachusetts }

\begin{abstract}
Using molecular dynamics simulations, we reveal emergent properties of hydrated electrode interfaces that while molecular in origin are integral to the behavior of the system across long times scales and large length scales. Specifically, we describe the impact of a disordered and slowly evolving adsorbed layer of water on the molecular structure and dynamics of the electrolyte solution adjacent to it. Generically, we find that densities and mobilities of both water and dissolved ions are spatially heterogeneous in the plane parallel to the electrode over nanosecond timescales. These and other recent results are analyzed in the context of available experimental literature from surface science and electrochemistry. We speculate on the implications of this emerging microscopic picture on the catalytic proficiency of hydrated electrodes, offering an new direction for study in heterogeneous catalysis at the nanoscale.    
\end{abstract}


\end{frontmatter}


\section{Introduction}	
\label{sec:intro}

Aqueous electrochemical interfaces have been the subject of renewed interest from the scientific community due to their role in sustainable energy technologies~\cite{Somorjai:2010p8235,Nrskov:2009p7956,brogioli2009extracting} such as batteries~\cite{aurbach2000review}, electrochemical capacitors~\cite{toupin2004charge,simon2008materials}, fuel cells\cite{turner1999realizable}, and electrocatalytic energy conversion and storage~\cite{lewis2006powering,khaselev1998monolithic,yan2013electrochemistry}. Recent progress in experimental surface science and molecular simulation have begun to expose, in unprecedented detail, the microscopic features of water-metal interfaces (See. Ref.~\cite{Carrasco:2012p3354} for a recent review). This attention has revealed that molecular correlations inherent to water-metal interfaces can give rise to complex emergent behavior with characteristic time and length scales that extend well beyond those which are assumed in traditional analytical theories. This observation has provoked a reexamination of the role of nanoscale fluctuations within this ubiquitous and often studied system\cite{spohr2003some,taylor2006first,guidelli2000recent,kornyshev2002electrochemical,wilhelm2010proton,spohr2002molecular,straus1995calculation,cao2014proton,Willard:2008p8256,limmer2013hydration,willard2013characterizing,limmer2013charge}. Here we extend this burgeoning perspective and highlight potential electrochemical implications using molecular dynamics simulations of the interface between a metallic electrode and an aqueous electrolyte. Specifically we show that nanoscale heterogeneity inherent to certain water-metal interfaces cause the densities and mobilities of both water and dissolved ions to deviate markedly from their in-plane averages. 

The current molecular understanding of water-metal interfaces derives primarily from experiments carried out at very low temperatures and ultra-high vacuum\cite{thiel1987interaction,henderson2002interaction}. These experiments combined with \textit{ab-initio} calculations\cite{Carrasco:2012p3354} have revealed that strong water-metal interactions can lead to the formation of an adsorbed monolayer of metal-bound water molecules. Individual adsorbed water molecules prefer configurations for which the oxygen atom is over the atop site of an exposed metal atom with molecular orientations that allow for the formation of favorable hydrogen bonds with neighboring members of the monolayer\cite{michaelides2003general}. At high surface coverage the monolayer patterns that emerge depend sensitively on a balance of water-metal adsorption energies, hydrogen bond energies, and the geometry of the exposed metal surface\cite{meng2004water}. This balance is reflected in the diversity of monolayer patterns that have been observed to form on different metal surfaces at low temperature\cite{Carrasco:2012p3354,Feibelman:2011p6303}.

At ambient conditions and when the metal is in contact with an aqueous phase, the ordered patterns of monolayer water molecules are expected to contain defects due to thermal fluctuations and to interactions with the bulk liquid. Unfortunately, under these conditions the molecular structure of the water-metal interface is not directly accessible to current experimental techniques and is outside of the reach of typical \textit{ab-initio} calculations. While indirect, molecular insight can still be gained through the analysis of experiments carried out at near-ambient conditions. For instance, temperature probed desorption has been used to examine the nature of the monolayer-ice interface~\cite{Kimmel:2005p8814}, where it was demonstrated that the adsorbed water monolayer is not ice-like, as previously postulated\cite{doering1982adsorption}, but in fact interacts weakly with subsequent layers of ice. These experiments are consistent with a molecular picture in which the adsorbed monolayer contains mostly in-plane hydrogen bonding, a particular water morphology that has been demonstrated to interact relatively weakly with subsequent layers of water~\cite{ogasawara2002structure,Kimmel:2005p8814,hodgson2009water,limmer2013hydration}. Electrochemical kinetic experiments and subsequent micro-kinetic modeling have produced a similar molecular picture. In particular for electrocatalytic oxygen reduction, it has been shown that an accurate modeling of the electrochemical current requires the inclusion of a phenomenological timescale\cite{hansen2014unifying} roughly equal to that found for the collective interfacial density fluctuations that arise due to weak interactions between the monolayer and the bulk liquid\cite{willard2013characterizing}. Further evidence for slow water dynamics has also emerged from experiments aimed at exploring electronic relaxation at the ice-metal interface~\cite{bovensiepen2008dynamic}. These experiments show that the timescales associated with electronic relaxation are many orders of magnitude larger at the ice-metal interface than at the vacuum-metal interface. This result may be similarly interpreted as  resulting from long correlation times for particular arrangements of adlayer water molecules that can serve as electronic trap sites. It has been suggested that such long correlation times for adlayer structure is a general feature of this type of interface~\cite{willard2013characterizing}.

\begin{figure}[t]
\begin{center}
\includegraphics[width=9cm]{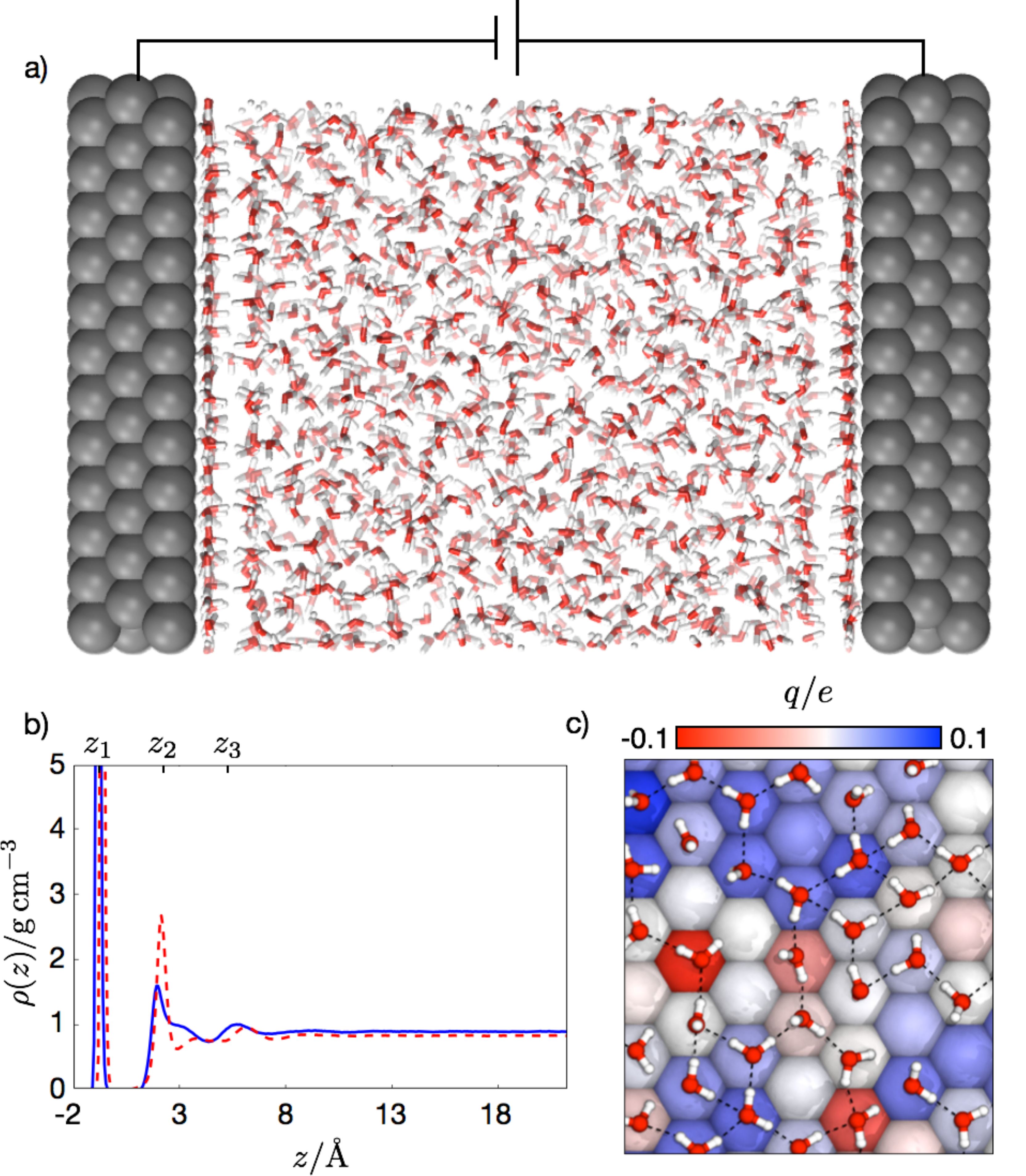}
\caption{Nanoscale simulation of aqueous metal interfaces. (a) Characetristic snapshot taken from the a molecular dynamics simulation of water between two constant potential platinum electrodes. (b) Density distribution for water exposed to the platinum 111 (red dashed lines ) or 100 (blue solid line) surface. (c) A view of the electrode surface and adsorbed monolayer in which electrode atoms have been colored to indicate their instantaneous partial charge.}
\label{Fi:1}
\end{center} 
\end{figure}

The affect of thermal disorder on the structure and dynamics of water-metal interfaces is difficult to predict due to the presence of collective molecular fluctuations that span a broad range of time and length scales. To describe the statistics of nanoscale fluctuations with molecular simulation requires a sacrifice of chemical accuracy in favor of computational efficiency. Here we utilize a class of molecular simulations\cite{reed2008electrochemical,limmer2013charge} whose quantum mechanical degrees of freedom are described implicitly, a simplification that provides access to time and length scales well beyond that which is available via more detailed calculations\cite{mattsson2003methanol,wang2004roles,SWG02}. The specific form of the interaction potentials and simulation methodology yield reasonable agreement with single molecular binding energy, ground state geometries\cite{Berendsen:1987p8660,Siepmann:1995p4868}, and accurately recover measured electrochemical properties like the potential of zero charge\cite{limmer2013hydration} and capacitance\cite{Willard:2008p8256,limmer2013charge} provided quantum corrections are taken into account. Specific details of the model and simulation algorithms are described in Refs.~\cite{Willard:2008p8256} and~\cite{limmer2013hydration}. Figure \ref{Fi:1}(a) illustrates the geometry of the system under investigation. The molecular nature of the system is manifest in the layering of the mean water density profile extending into the bulk from the electrode as shown in Fig.~\ref{Fi:1}(b), and in an example of the instantaneous charge distribution generated from the polarizable metal electrode as shown in Fig.~\ref{Fi:1}(c), neither of which would emerge in the context of traditional mean field theories. 

These molecular simulations predict that the monolayer contains disordered patterns of water molecules that equilibrate on time scales that are several orders of magnitude larger than that of the bulk liquid and in a manner which is spatially heterogeneous and governed locally by defects in the two-dimensional hydrogen bond network. This molecular picture deviates from standard mean-field descriptions of electrochemical interfaces that assume a homogeneous and swiftly relaxing medium. Such mean-field theories originated from the microscopic model of Guoy and Stern and were postulated largely on thermodynamic principles, characterizing the interface as a dielectric continuum with a depressed polarizability near the electrode\cite{trasatti1981effect,kolb2002atomistic,bonthuis2011dielectric} and in some instances a complex frequency dependence in its response\cite{green2002fluid,levitan2005experimental}. While such models provide a means for interpreting electrocapilary and cyclic voltametry measurements, especially at low to moderate electrolyte concentration\cite{bagotsky2005fundamentals}, their neglect of molecular detail render them incapable to explaining the origin of the emergent interfacial properties and insensitive to many microscopic features pertinent to catalysis. For instance, short timescale solvent dynamics and specific ion effects are well known to effect electrochemical reaction rates\cite{hansen2014unifying,anson1975patterns}, yet are difficult to explain without a molecular-level description of the system. 

We have separated our results into two sections. First, in Section~\ref{Se:water} we quantify the structural and dynamic properties of the interface between the electrode adsorbed water monolayer and bulk liquid water. We illustrate that over nanosecond timescale the region of bulk liquid that is directly adjacent to the adsorbed water monolayer is both structurally and dynamically heterogeneous in response to the structure and dynamics of the underlying water monolayer. Second, in Section~\ref{Se:ion} we demonstrate that for a suite of aqueous ions, the spatial distribution and the dynamics are similarly heterogeneous. These findings suggest that monolayer disorder likely plays an important role in governing the chemical kinetics of charge redistribution and potentially reactivity. 

\section{Monolayer Heterogeneity is Reflected in the Bulk Liquid Interface}\label{Se:water}
In this section we quantify the influence of monolayer heterogeneity on the structure and dynamics of the adjacent liquid by considering the average density and translational mobility of water near the adsorbed monolayer. The surface geometry of the electrode, the 111 face of platinum for this study, provides a template for the positions of water molecules within the adsorbed water monolayer. Although the electrode presents a perfectly ordered surface, at ambient conditions the monolayer itself is not perfectly ordered and contains defects in hydrogen bonding (i.e. molecular orientation) as well as an equilibrium concentration of vacancies in the form of unoccupied surface metal atoms. 

\begin{figure}[fh!]
\begin{center}
\includegraphics[width=8.7cm]{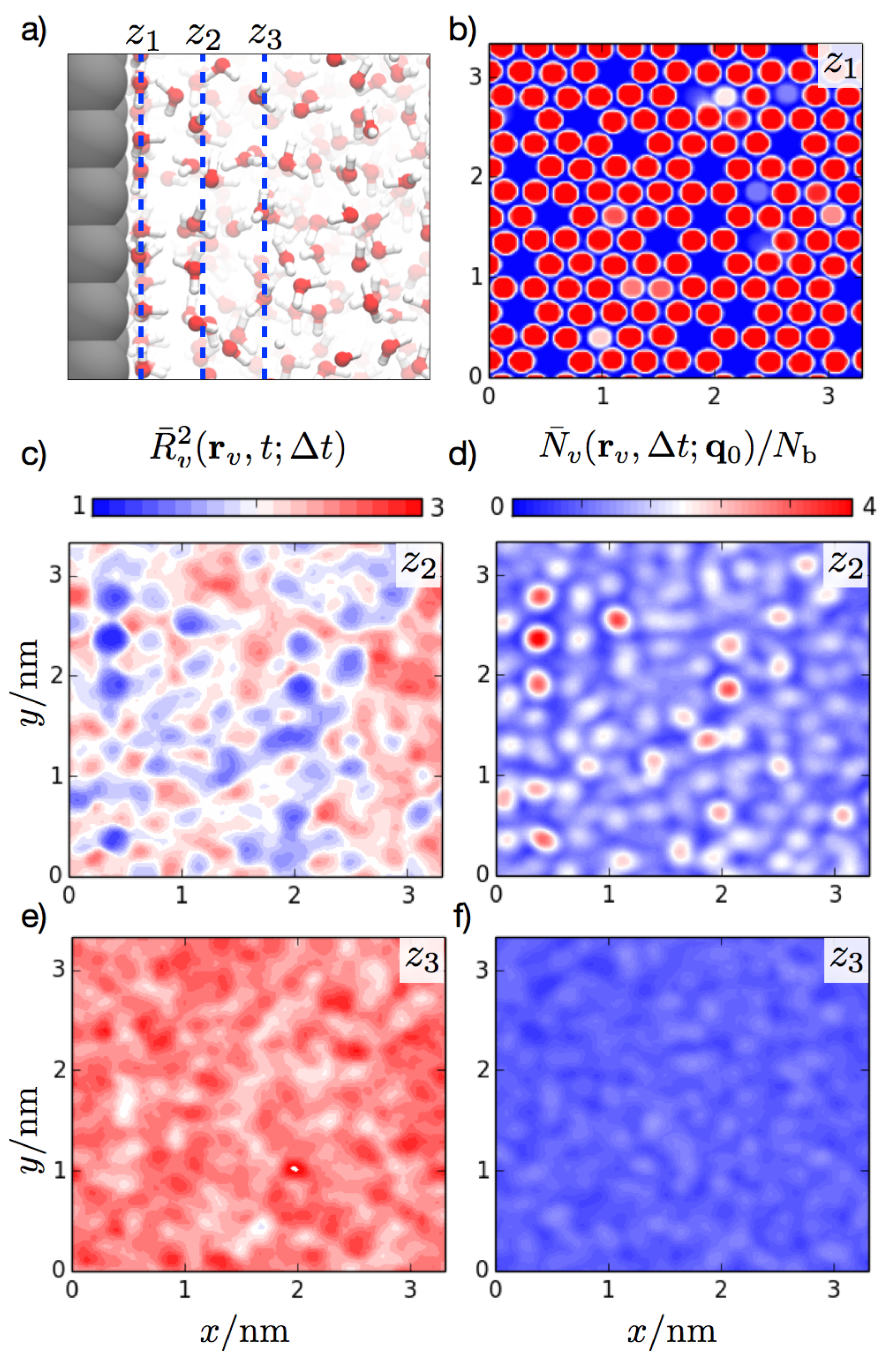}
\caption{(a) A snapshot of the water-metal interface. The three $z-$values indicated, $z_1$, $z_2$, and $z_3$ correspond to plane of the adsorbed water monolayer, the first peak in the bulk liquid density profile, and the first plateau in the bulk liquid density profile respectively. The contour maps in panels (b), (d), and (f) correspond to the average solvent density, $\bar{N}_v(\mathbf{r}_v,\Delta t;\mathbf{q}_0)$ in the plane parallel to the metal surface at vertical position equal to $z_1$, $z_2$, and $z_3$ respectively. Each panel has been normalized by $N_\mathrm{b}$, the average value of $\bar{N}_v(\mathbf{r}_v,t;\Delta t)$ for $\mathbf{r}_v$ in the bulk liquid. The contour maps in panels (c) and (e) correspond to the average translational mobility, $\bar{R}_v(\mathbf{r},t;\Delta t)$, for water molecules originating at a specific position in the plane of the electrode and with a $z$ value equal to $z_2$ and $z_3$ respectively. Panels b-f are all averaged over an identical $\Delta t$ length segment of the simulation data.}
\label{Fi:3}
\end{center} 
\end{figure}

The disordered water monolayer is physically distinct from the bulk liquid, they are separated by a region of near-zero water density as can be seen in Fig.~\ref{Fi:1}(b) and Fig.~\ref{Fi:3}(a) and their equilibrium characteristics are governed by very different timescales~\cite{willard2013characterizing}.  The timescale over which correlations in molecular orientations decay are several orders of magnitude larger for molecules within the adsorbed monolayer than for molecules within the bulk. We denote their average relaxation times as $\tau_\mathrm{S}$ and $\tau_\mathrm{B}$ respectively. Fluctuations within the monolayer and the adjacent bulk liquid are coupled by collective fluctuations of the bulk liquid interface, which themselves are characterized on timescales intermediate to the monolayer and the bulk~\cite{limmer2013hydration}. Over the picosecond timescales that are characteristic of molecular fluctuations in the bulk, the structure of the adsorbed water monolayer is essentially fixed and the transient spatial disorder within the adlayer is reflected in the structure and dynamics of the adjacent bulk liquid interface. In this way heterogeneity within the adsorbed monolayer can exert an indirect influence on the dynamics of those electrochemical processes that couple to the dynamics of the solvent, such as proton transfer which is modulated by fluctuations in solvent polarization.  

The influence of monolayer disorder on the molecules within the adjacent bulk liquid can be seen by analyzing the spatial dependence of the water density in the vicinity of the electrode. Specifically we compute the water density within a spherical probe volume $v$ centered at a fixed position $\mathbf{r}_v$, and averaged over a time $\Delta t$. Formally we compute,
\begin{equation}
\bar{N}_v(\mathbf{r}_v,\Delta t;\mathbf{q}_0)=\frac{1}{\Delta t} \int_0^{\Delta t} dt' \left ( \sum_{i=1}^{N_\mathrm{w}} \Theta(\vert \mathbf{r}_i(t')-\mathbf{r}_v) \vert - \sigma_v) \right ),  
\label{Eq:1}
\end{equation}
where $\tau_\mathrm{b} << \Delta t < \tau_\mathrm{s}$~\cite{timescales}), the summation is over all $N_\mathrm{w}$ water molecules, $\mathbf{r}_i(t)$ is the position of the oxygen of the $i$th water molecule at time $t$, $\mathbf{r}_v$ is the fixed position of the spherical probe volume of radius $\sigma_v=1\mathrm{\AA}$, and $\Theta(x)$ is a step function equal to unity when $x\le 0$ and equal to zero when $x>0$. For the averaging time we have chosen, i.e. $\Delta t = 1$ns, the system cannot adequately sample monolayer disorder and therefore the average in Eq.~\ref{Eq:1} reflects a parametric dependence on the initial configuration of the adsorbed water monolayer, denoted by $\mathbf{q_0}$. 

Figures~\ref{Fi:3}(b), (d), and (f) contain plots of the interfacial density field of Eq.~\ref{Eq:1}, resolved along the plane parallel to the electrode surface at three different distances from the surface of the electrode, each generated by computing $\bar{N}_v(\mathbf{r}_v,t;\Delta t)$ for the manifold of probe volume positions with fixed $t$ and $z$ (i.e. the coordinate perpendicular to the exposed electrode surface). As indicated in Fig.~\ref{Fi:1}(b) and Fig.~\ref{Fi:3}(a) the planes defined by $z_1$, $z_2$, and $z_3$ correspond to the position of the adsorbed water monolayer, the first peak in bulk liquid density profile, and the first plateau in the bulk liquid density profile respectively. 

As seen in Fig.~\ref{Fi:3}(b) the surface geometry of the metal electrode is clearly evident from the density profile of the adsorbed water monolayer, and the specific pattern of vacancies persist over the nanosecond averaging. The monolayer heterogeneity is reflected in the density profile of the bulk liquid interface (i.e., at $z=z_2$) in the form of a pattern of \textit{hot spots} of various amplitude, some as high as five times that of the average bulk liquid. The correlation between the positions of the monolayer vacancies (low density regions in Fig.~\ref{Fi:3}(b)) and the high density regions of the bulk liquid interface (hot spots in Fig.~\ref{Fi:3}(d)) indicate that molecules in the bulk prefer to nestle into the holes presented by surface vacancies. However the spatial variations in amplitude of $\bar{N}_v(\mathbf{r}_v,\Delta t;\mathbf{q}_0 )$ at $z_2$ indicates a complex coupling between monolayer disorder and the adjacent bulk liquid. Although spatial variations in the solvent density at $z=z_2$ persist on timescales well beyond that of typical bulk density fluctuations i.e., $\Delta t >> t_\mathrm{b}$, one cannot predict the data shown in Fig.~\ref{Fi:3}(b) from the data shown in Fig.~\ref{Fi:3}(d) alone. The position $z_3$ is approximately one molecular diameter further into the bulk liquid than $z_2$ and at that distance density fluctuations are spatially homogeneous, distributed narrowly around a value which is very near that of the average bulk density. This indicates the influence of monolayer heterogeneity does not extend far into the bulk, and that density fluctuations perpendicular to the electrode are characterized by a correlation length of a few molecular diameters.

The dynamics of individual water molecules at the bulk liquid interface are also sensitive to the monolayer-induced spatial heterogeneity. A spatially local measure of translational mobility, which like Eq.~\ref{Eq:1} depends parametrically on $\mathbf{q}_0$, is given by,
\begin{equation}
\bar{R}^2_v(\mathbf{r}_v,\Delta t;\mathbf{q}_0) = \frac{\int_0^{\Delta t} dt' I (\mathbf{r}_v,t';\mathbf{q}_0)}{\Delta t \bar{N}_v(\mathbf{r}_v,\Delta t;\mathbf{q}_0)}
\label{Eq:2}
\end{equation}
where,
$$
I(\mathbf{r}_v,t;\mathbf{q}_0) = \sum_{i=1}^{N_\mathbf{w}}  \vert \mathbf{r}_i(t) - \mathbf{r}_i(t+\delta t) \vert^2 \Theta(\vert \mathbf{r}_i(t) - \mathbf{r}_v \vert - \sigma_v )
$$
is the mean squared displacement for molecules originating within a probe volume $v$ within the time window extending between $t$ and $t+\Delta t$. Here the time increment $\delta t=10$ps has been chosen to be equal to the time taken for a water molecule to diffuse a single molecular diameter within the bulk liquid. Figure~\ref{Fi:3}(c) and (d) shows $\bar{R}^2_v(\mathbf{r}_v,t;\Delta t)$ computed for the set of probe volumes positioned at fixed $t$ and $z_2$ and $z_3$ respectively (identical to those used above to compute $\bar{N}_v(\mathbf{r}_v,t;\Delta t)$).  Translational mobility within the adlayer is effectively nonexistent, and occurs only after rare desorption events that destabilize local hydrogen bonding patterns.

We observe a significant spatial dependence in the average translational mobility the details of which bear resemblance to the spatial patterns observed in $\bar{N}_v(\mathbf{r}_v,\Delta t;\mathbf{q}_0)$ computed at the same value of $z$. This indicates a  coupling between the average structure and microscopic dynamics at the water-metal interface. Generally, regions of particularly low mobility correspond to regions of high density, a signature of the pinning of the molecule above surface vacancies. Exploring how these spatially dependent interfacial properties affect electrolyte species is the focus of the next section. 

\section{Diffusion and Distributions of Electrolytes}
\label{Se:ion}
We consider four different monovalent non-polarizable ions, $\mathrm{Cl}^-$, $\mathrm{I}^-$, $\mathrm{Na}^+$, and $\mathrm{Cs}^+$. The ion-ion and ion-water interaction parameters are taken from Ref.~\cite{koneshan1998solvent}. The ion-metal potentials are adapted form Ref.~\cite{Willard:2008p8256}, though in what follows we explicitly consider only those properties of the ions away from the metal surface as effects from charge transfer are likely important at ion-metal contact. Using this model, we demonstrate below that the static and dynamic properties of ions at the bulk water interface exhibit long-lived spatially dependent heterogeneities. To begin we consider the diffusion of ions located on the bulk side of the bulk-monolayer interface.
\begin{figure}[t]
\begin{center}
\includegraphics[width=9cm]{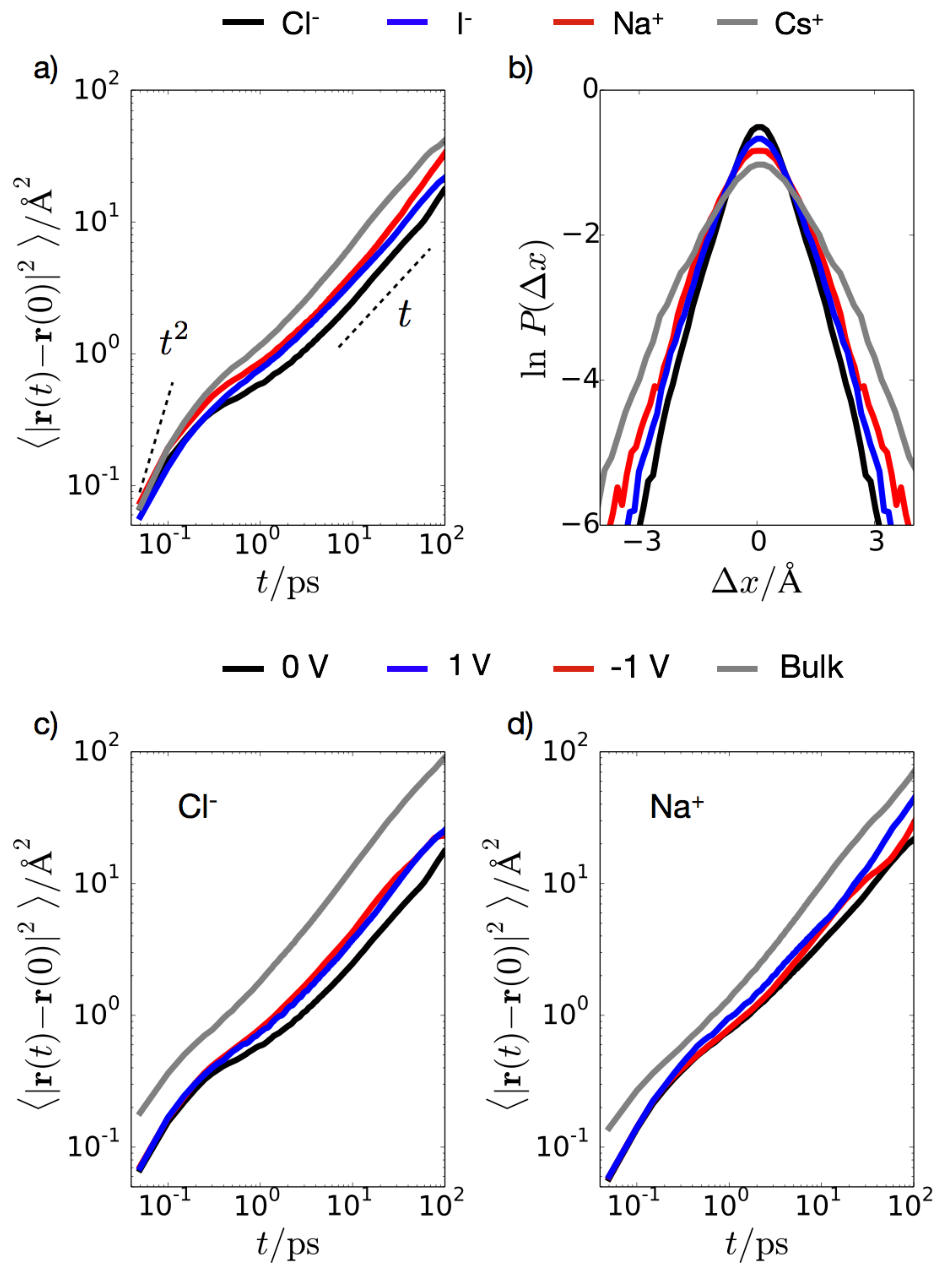}
\caption{Transport properties of individual ions at the bulk-monolayer interface. Panel (a) shows mean squared displacement and panel (b) shows the distribution of lateral displacements. In Panel (a), dashed lines of constant slope 1 and 2 illustrating the crossover from ballistic to diffusive motion. Panels (c) and (d) show the mobility of Cl$^-$ and Na$^+$, respectively, under applied potential and in the bulk.}
\label{Fi:4}
\end{center} 
\end{figure}

\subsection{Correlated ionic mobility at the aqueous interface}
Near the interface, the diffusion of individual ions is correlated with the slow relaxation of the adsorbed water layer. Figure~\ref{Fi:4}(a) shows the mean-squared displacement, $\langle \vert \mathbf{r}(0) - \mathbf{r}(t) \vert^2 \rangle$ for individual ions whose initial position $\mathbf{r}(0)$ is within 5$\mathrm{\AA}$ of the adsorbed water monolayer. The qualitative behavior, which is shared by all four species of ions under consideration, shows the expected crossover from ballistic motion at short times, $\sim t^2$, to diffusion motion at long times, $\sim t$. Between these two behaviors there is a plateau where their mean square displacement is nearly constant.  The extent of this plateau depends on the identity of the ion, and occurs between 0.1 and 2 picoseconds.  This plateau does not occur for diffusion in the bulk  at these conditions.

The plateau, or caging, is reminiscent of diffusion in supercooled liquid\cite{kob1995testing} and coorrespondingly ionic diffusion at the interface reflects the dynamic heterogeneity expressed by the adsorbed monolayer. Molecularly, this caging behavior results from the occasional \textit{sticking} of an ion to the surface of the monolayer that occurs when a member of the electrode-adsorbed monolayer participates in the ion's solvation shell. When this happens the diffusion of an ion is coupled to the slow reorientation of the monolayer, until a fluctuation occurs that exchanges the electrode adsorbed member of the solvation shell for a more mobile molecule, either on the surface or within the bulk solvent. Larger ions, which are more weakly hydrated, are less sensitive to individual slow adsorbed water molecules, showing the weakest plateau. By contrast smaller anions, whose motion is the most correlated with donated hydrogen bonds, are the most sensitive, showing the largest plateau time.  

Figure~\ref{Fi:4}(b) plots $p(\Delta x)$, the probability for an ion to make a lateral displacement $\Delta x$,  over a time $t_\mathrm{d}$ given by,
\begin{equation}
p(\Delta x) = \frac{\left \langle \delta( x(0) - x(t_\mathrm{d}) - \Delta x) \Theta(z(0) - z^*) \right \rangle}{\left \langle \Theta(z(0) - z^*) \right \rangle},
\end{equation}
where $x(t)$ and $z(t)$ are the $x$ and $z$ Cartesian coordinates respectively for the ion at time $t$, $\delta(x)$ is the Dirac delta function. To reflect diffusive motion, we take $t_\mathrm{d}=10$ps, which is longer than the plateau period for all ions. At these timescales, the statistics governing lateral displacements have pronounced non-Gaussian tails, indicating that large displacements are more probable than one would estimate based on the mean behavior. There is a clear dependence on ion type, with negative ions making fewer large displacement than similarly sized positive ions. Similarly to the mean squared displacement, these features are reminiscent of that found in a supercooled liquids and by analogy reflect the correlated dynamics of the frustrated adlayer of water. 

We have also evaluated the mean squared displacement of Cl$^-$ and Na$^+$ under a constant applied electric potential and away from the interface at 0 applied potential. The thermodynamic state of the system is determined by the potential difference across the simulated electrochemical cell. In order to avoid an ambiguity that arises do to the inversion symmetry of our simulation cell, we define the applied potential as $\Delta \Psi = \Psi_{L} - \Psi_{R}$ where $\Psi_{L}$ is the potential of the left electrode near $z=0$ and $\Psi_{R}$ is the potential of the right electrode at large $z$.  The curves shown in Figs.~\ref{Fi:4}(c-d) reflect properties near the left electrode. For all cases the long time behavior yields a diffusion constant, which is depressed compared to the bulk value. For $\mathrm{Cl}^-$ the diffusion constant is a factor of 8 smaller at the interface compared to the bulk, while for $\mathrm{Na}^+$  the diffusion constant  is a factor of 4 smaller at the interface compared to the bulk. As shown in Fig.~\ref{Fi:4}(c,d), not only is the magnitude of the displacement large, but the plateau behavior resulting from the correlations within the slow surface water is absent. For both ions under applied potential the ionic mobility in the plane of the electrode is enhanced, which is a consequence of the reorientation of the surface water at the electrode\cite{willard2013characterizing}. 
\begin{figure}[fh!]
\begin{center}
\includegraphics[width=8.5cm]{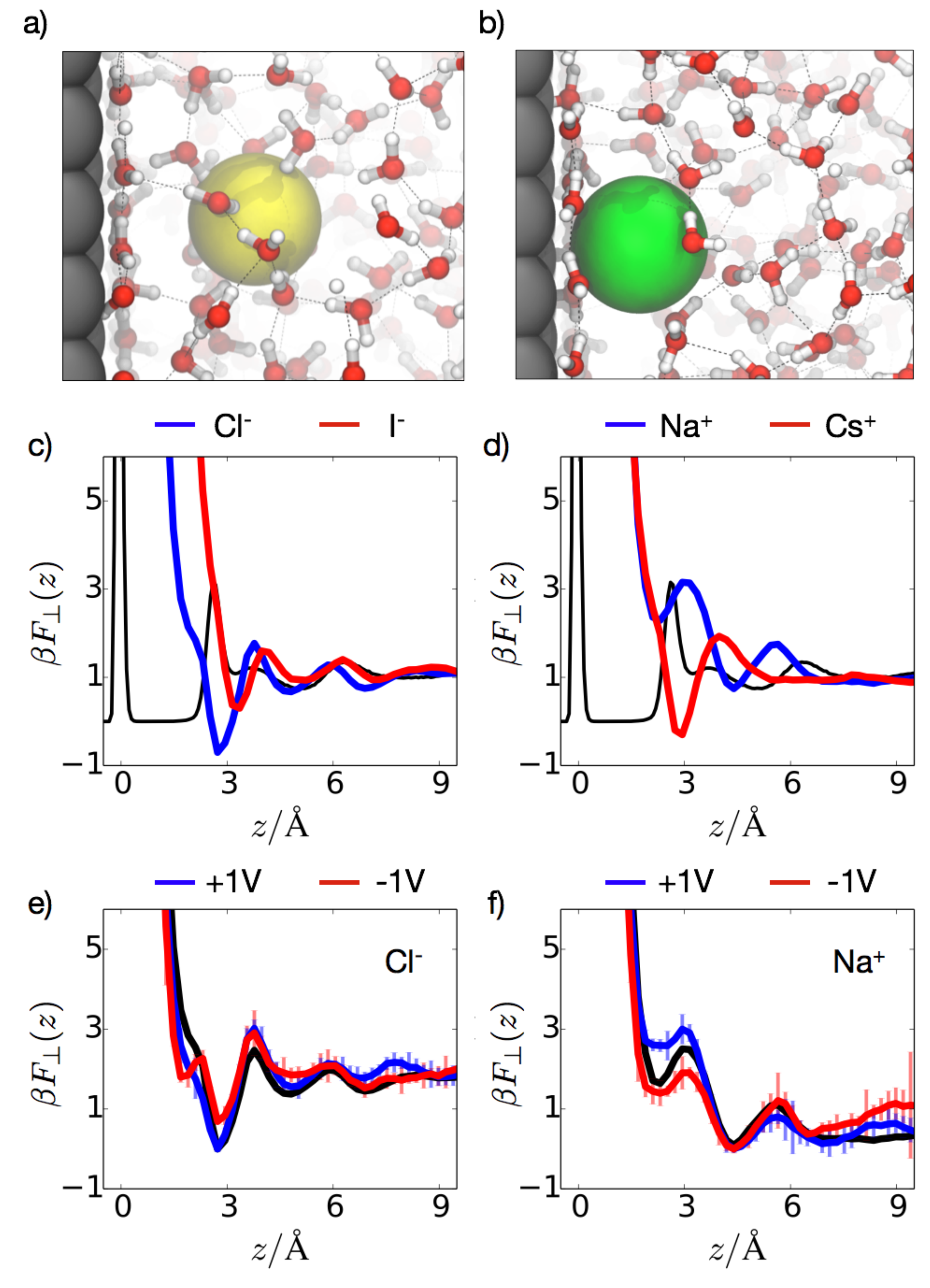}
\caption{Dependence on ion type and applied potential on the distribution of ions away from the electrode interface. (a-b) Representative configurations of an (a) Cl anion and (b) Na cation rendered to scale with plots in (c-f). (c-d) Free energies as a function of $z$ for 2 anions and 2 cations, with the water density distribution shown in black. (e-f) Free energies as a function of $z$ away from an electrode under different applied potential for the Cl$^-$ anion and Na$^+$ cation. For reference, the black curves in (e-f) show the results for 0 applied potential.}
\label{Fi:5}
\end{center} 
\end{figure}

\subsection{Specific ion adsorption at the aqueous interface}
The partitioning of ions between the bulk liquid and the interface and also laterally within the monolayer-bulk interface depends on ion type, electrode potential, and exhibits long-lived spatial heterogeneity. Using umbrella sampling\cite{frenkel2001understanding} we have obtained an accurate estimate of the relative positions of ions. Specifically we have computed the potential of mean force resolved in the direction perpendicular to the electrode surface given by, 
\begin{equation}
\beta F_\perp(z) = -\ln [\langle \delta(z'-z) \rangle],
\end{equation}
\begin{figure*}
\begin{center}
\includegraphics[width=17cm]{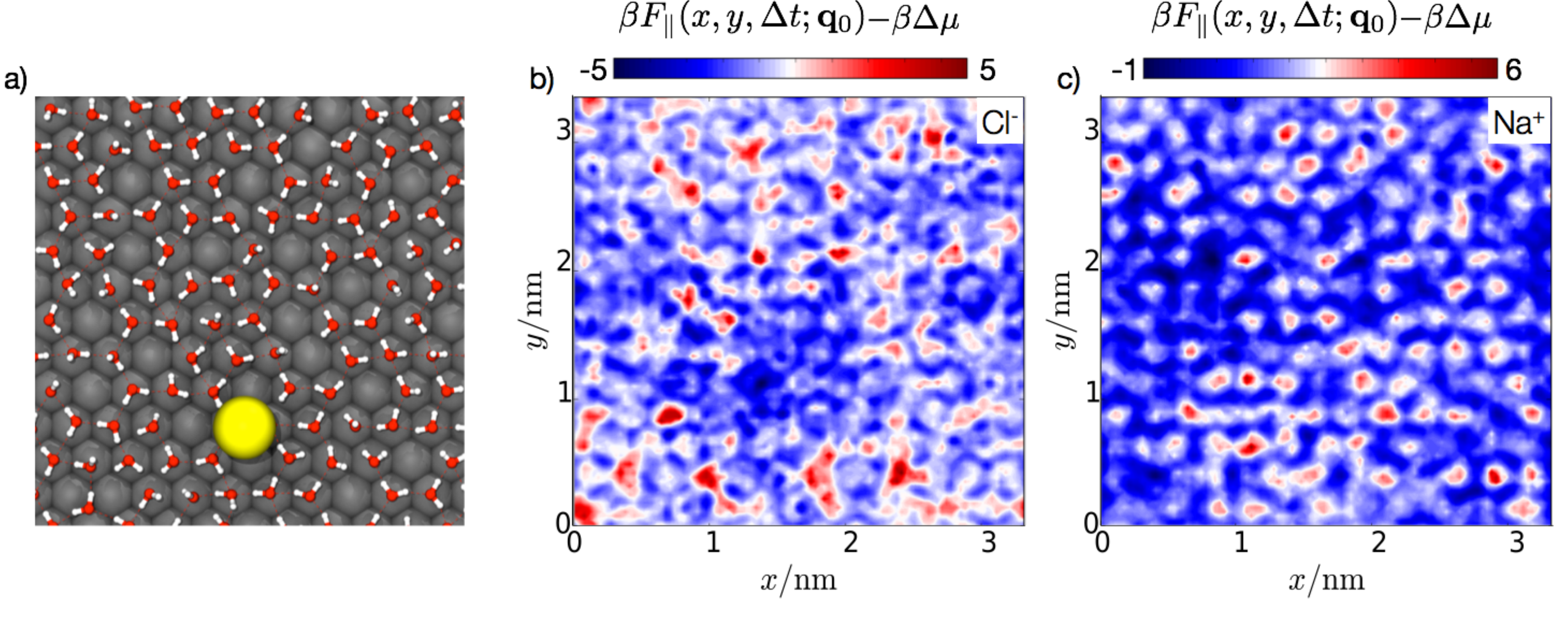}
\caption{Transient ionic solvation properties in the plane of the electrode. (a) Representative configuration of an ion (yellow) in contact with the water adlayer. Surrounding liquid water are not shown for clarity. (b) Time dependent solvation potential in for Cl$^-$ averaged with $\Delta t=$0.5 ns.  (c) Time dependent solvation potential in for Na$^+$ with $\Delta t=$0.5 ns. }
\label{Fi:6}
\end{center} 
\end{figure*}

where $\beta=1/k_\mathrm{B} T$, $k_\mathrm{B}$ is the Boltzmann constant, $T$ is temperature, and $z'$ denotes the dynamic $z$ position for the ionic species under consideration. Results for the four ions studied in the previous sections are shown in Fig.~\ref{Fi:5}(c-f). The molecular level detail provided by these simulations is evident in the explicit layering within the free energy away from the electrode, as also seen in the density profile in Fig.~\ref{Fi:1}. This layering arises as a consequence of molecular packing and yields a qualitatively different charge density profile from the exponentially decreasing profile expected from Gouy-Chappman theory or modified Poission-Boltzmann approaches\cite{bagotsky2005fundamentals}. The sharp rise in all of the curves upon approaching the water adlayer results from the large free energy energy required to displace single water molecules on the metal surface. With an accurate representation of the metal-ion force, the distribution of some of these ions at the metal surface would likely show signs of contact adsorption with an accompanying deep minima in the free energy at  $z\approx 0$\cite{anson1975patterns}. Without such a description, we explicitly only consider the free energy for values of $z$ past the first water adlayer.

The partitioning of ionic species between the interface and the bulk depends sensitively on molecular details. Figure~\ref{Fi:5}(c) shows that large anions have a preference for the bulk liquid interface, though in this case Cl$^-$ is more strongly attracted to the interface then I$^-$. The stronger adsorption of Cl$^-$ over I$^-$ is due to the fact the former can approach vacant sites on the electrode surface while the latter is too large to sit comfortably within such sites, which are typically coordinated by 6 waters. The tendency of Cl$^-$ to occupy regions of the liquid interface directly adjacent to monolayer vacancies indicates that the overall concentration of interfacial Cl$^-$ is likely quite sensitive to details effecting monolayer structure such as electrode geometry. Figure~\ref{Fi:5}(d) shows that a large cation, Cs$^+$ has a preference for the bulk liquid interface, while an strongly hydrated smaller cation, Na$^+$ is not selectively surface adsorbed. These behaviors agree with previous experimental adsorption measurements\cite{anson1975patterns}, and follow roughly the same trends seen in the Hoffmister series\cite{jungwirth2008ions}. 

We have additionally computed $F_\perp(z)$ for $\mathrm{Na}^+$ and $\mathrm{Cl}^-$ for applied electrode potentials of $\pm 1V$. The results are plotted in Fig.~\ref{Fi:5}(e-f). The changes in $F_\perp(z)$ with applied electrode potential are subtle, generating like-charged repulsion (positive electrode and $\mathrm{Na}^+$ or negative electrode and $\mathrm{Cl}^-$) is more evident than generating oppositely charge attraction. The weak dependence on potential close to the electrode reflects the effectiveness with which the first adsorbed layer of water screens the applied potential\cite{Willard:2008p8256,limmer2013charge}. 

As discussed previously,  the separation of timescales between surface and bulk relaxation, means that on intermediate timescales, $\tau_\mathrm{B} \ll t < \tau_\mathrm{s} $, a given surface configuration $\mathbf{q}_0$ at time $t=0$ templates a laterally heterogeneous solvation potential. This potential is defined by,
\begin{equation}
\beta F_\parallel(x,y,\Delta t;\mathbf{q}_0) \\
= - \ln \Big [ \frac{1}{\Delta t} \int_0^{\Delta t} \,dt' \, G(x,y,t';\mathbf{q}_0) \Big ]
\label{Eq:Pv_t_r}
\end{equation}
where 
$$
G(x,y,t;\mathbf{q}_0) =\delta(x'(t';\mathbf{q}_0)-x) \delta(y'(t';\mathbf{q}_0)-y) \Theta(z'(t;\mathbf{q}_0)-z)\\
$$
and $t$ is the timescale over which the distribution is averaged, $x'$ and $y'$ are the instantaneous $x$ and $y$ positions for the ion species under consideration and $z=3$ to constrain statistics to the bulk-liquid interface. The fluctuations of the dynamic variables,  $x'$ and $y'$, over these short timescales depend parametrically on the initial surface condition. 

Figure~\ref{Fi:5} shows $\beta F_\parallel(x,y,\Delta t;\mathbf{q}_0)$ for $\mathrm{Na}^+$ and $\mathrm{Cl}^-$ at zero applied electrode potential and $t=0.5 \mathrm{ns}$, relative to the average free energy difference for bringing the ion from the bulk to the interface, $\beta \Delta \mu = \beta F_\perp(9)- \beta F_\perp(3)$, as computed from Fig.~\ref{Fi:4}. Similar to those results plotted Fig.~\ref{Fi:3}(b-d), $\beta F_\parallel(x,y,\Delta t;\mathbf{q}_0)$ exhibits long-lived spatial heterogeneities that vary over several times the thermal energy, $k_\mathrm{B}T$. The disorder contains spatial patterns that reflect that of the adsorbed water monolayer (and thus indirectly the electrode surface itself). Focusing on $\beta F_\parallel(x,y,\Delta t;\mathbf{q}_0)$ for $\mathrm{Na}^+$, we observe that despite an overall average preference for bulk solvation, as demonstrated through $\beta F_\perp(z)$, certain specific lateral regions are significantly attractive to $\mathrm{Na}^+$ relative to the bulk. The patterns in the distribution persist on timescales long relative to ion diffusion meaning that for an ion either near or within the bulk liquid interface there exist persistent driving forces of several times $k_\mathrm{B}T$ pushing ions towards or away from specific regions of the bulk liquid interface. 

\section{Perspectives for future work}
In this work, we have illustrated a number of consequences that a slowly-evolving, disordered adlayed of water on a metal surface has on the equilibrium fluctuations and dynamics of a contact electrolyte solution. While the text focused selectively on the 111 crystal face of platinum, previous work has shown that qualitative features of the interface are largely conserved in other systems, provided the metal surface strongly binds water and enforces a predominately two-dimensional hydrogen-bonding network\cite{Willard:2008p8256}. 

An important future direction for this work is to elucidate the kinetic implications of these altered solvation properties at the interface, specifically in regards to heterogeneous catalysis. Along these directions, the results of Ref.~\cite{hansen2014unifying} are an important step forward. The development of simulation methodologies that allow for select quantum degrees of freedom to be treated while maintaining the computational efficiency of the predominantly classical molecular dynamics system will be instrumental in this pursuit at molecular scales. References \cite{cao2014proton,golze2013simulation} are initial steps in this direction.  In the shorter term, studies elucidating the role of the slowly evolving adlayer on reactant adsorption on the metal or on charge separation or recombination near the bulk liquid interface will likely already aid in the understanding simple reactions. These and similar studies are underway.

\section*{Acknowledgment}
This research used resources of the National Energy Research Scientific Computing Center, a DOE Office of Science User Facility supported by the Office of Science of the U.S. Department of Energy under Contract No. DE-AC02-05CH11231. DTL is supported as a fellow of the Princeton Center for Theoretical Science. 


\bibliographystyle{elsarticle-num}



\end{document}